\documentclass[aps,twocolumn,floatfix, showpacs, superscriptaddress]{revtex4-1}
\usepackage[pdftex]{graphicx}
 \usepackage{amsmath}
\usepackage[T1]{fontenc}
\usepackage{amssymb}
\usepackage{amsfonts}
\usepackage[breaklinks=true, colorlinks=true, linkcolor= blue, urlcolor= blue, citecolor= blue]{hyperref}
\usepackage{bm}
 \usepackage{amsmath} 
  \usepackage{flushend}

\usepackage{color}
\usepackage{lipsum}
\usepackage{amsfonts} 
\usepackage{amssymb, mathrsfs}
\usepackage{braket}
\usepackage{graphicx} 
\usepackage{subfigure}
\usepackage{bbm}
\def\beq{\begin{equation}}
\def\eeq{\end{equation}}
\def\bsp{\begin{split}}
\def\esp{\end{split}}
\def\bea{\begin{eqnarray}}
\def\eea{\end{eqnarray}}
\def\ba{\begin{array}}
\def\ea{\end{array}}

\def\dg{\dagger}

\def\lb{\left(}
\def\rb{\right)}

\def\l.{\left.}
\def\r.{\right.}

\def\ra{\rangle}
\def\la{\langle}

\def\bo{{\vec k}}

\begin{document}

\title{Photo-induced Floquet Weyl magnons in  noncollinear antiferromagnets}
\author{S. A. Owerre}
\affiliation{Perimeter Institute for Theoretical Physics, 31 Caroline St. N., Waterloo, Ontario N2L 2Y5, Canada.}

\begin{abstract}
We study periodically driven insulating noncollinear stacked kagome antiferromagnets with a conventional symmetry-protected three-dimensional (3D) in-plane $120^\circ$  spin structure, with either positive  or negative vector chirality. We show that the symmetry protection of the  in-plane $120^\circ$  spin structure can be broken in the presence of an off-resonant circularly or linearly polarized electric field propagating parallel to the in-plane $120^\circ$  spin structure (say along the $x$ direction). Consequently,  topological Floquet Weyl magnon nodes with opposite chirality are photoinduced along the $k_x$ momentum direction. They manifest as the monopoles of the photoinduced Berry curvature.  We also show that the system exhibits a photoinduced magnon thermal Hall effect for circularly polarized electric field. Furthermore, we show that the photoinduced chiral spin structure is a canted 3D in-plane $120^\circ$  spin structure, which  was recently observed in the equilibrium noncollinear antiferromagnetic Weyl semimetals Mn$_3$Sn\slash Ge.   Our result not only paves the way towards the experimental realization of Weyl magnons and photoinduced thermal Hall effects, but also provides a powerful mechanism for manipulating the intrinsic properties of 3D topological antiferromagnets. 

\end{abstract}

\maketitle

\section{Introduction}
In recent years, the Floquet engineering of topologically nontrivial systems from topologically trivial ones has attracted considerable attention in electronic systems  \cite{foot3,foot4,lin,foot5,gru,fot,jot,fla,we2,we4,we5, gol,buk,eck,du,du1,du2,tak,mat, zyan, hhbe, rwang,ste}. However, its application to realistic three-dimensional (3D) materials is very limited.  To our knowledge, a photoinduced Floquet Weyl semimetal has only been created in the non-magnetic 3D Dirac semimetal Na$_3$Bi \cite{hhbe}. In the electronic Floquet topological systems, the intensity of light is characterized by a dimensionless quantity given by \cite{foot3,foot4,lin,foot5,fot}

 \bea 
 \mathcal E_i = \frac{e E_i a}{\hbar \omega},
 \label{Fp}
 \eea 
 where $E_i$ ($i= x,y,z$) are the amplitudes of the electric  field, $e$ is the electron charge, $\omega$ is the angular frequency of light, $a$ is the lattice constant, and $\hbar$ is the reduced Plank's constant.  In the magnetic bosonic Floquet topological systems, however, the intensity of light is characterized by a dimensionless quantity  given by \cite{sol0}
 
  \bea 
 \mathcal E_i = \frac{g\mu_B E_i a}{\hbar c^2},
 \label{eq1}
 \eea 
 where  $g$ is the Land\'e g-factor, $\mu_B$ is the Bohr magneton, and $c$ is the speed of light. The lack of frequency dependence in Eq.~\eqref{eq1} unlike in  Eq.~\eqref{Fp}  is because it originates  from the  Aharonov-Casher phase \cite{aha}, which is a topological phase acquires by a neutral particle with magnetic dipole moment (such as magnon) in the presence of an electric field.   By equating the two  dimensionless quantities we can see that  the spin magnetic dipole moment $ g\mu_B$ carried by magnon in the periodically driven insulating magnets is given by 
   \bea 
 g\mu_B = \frac{ec^2}{\omega} =\frac{ec\lambda}{2\pi}.
 \label{eq3}
 \eea 
Therefore for an experimentally feasible light wavelength $\lambda$ of order $10^{-8}$m, the spin magnetic dipole moment $ g\mu_B$ carried by magnon in the periodically driven insulating magnets  is comparable to the electron charge $e$. This shows that the  electronic Floquet topological systems \cite{foot3,foot4,lin,foot5,fot} are indeed similar to the magnetic bosonic Floquet topological systems recently studied in insulating ferromagnets with finite magnetization \cite{sol0,sol1,sol2,sol3}. In magnetic systems, however, the Floquet physics can reshape the underlying spin Hamiltonian to stabilize magnetic phases  and provides a promising avenue for  inducing and tuning topological spin excitations in trivial quantum magnets \cite{sol0,sol1,sol3}, as well as photoinduced topological phase transitions in intrinsic topological magnon insulators \cite{sol2}.  This formalism  provides a direct avenue  of generating and manipulating ultrafast spin current using terahertz radiation \cite{walo}. 
\begin{figure}
\centering
\includegraphics[width=1\linewidth]{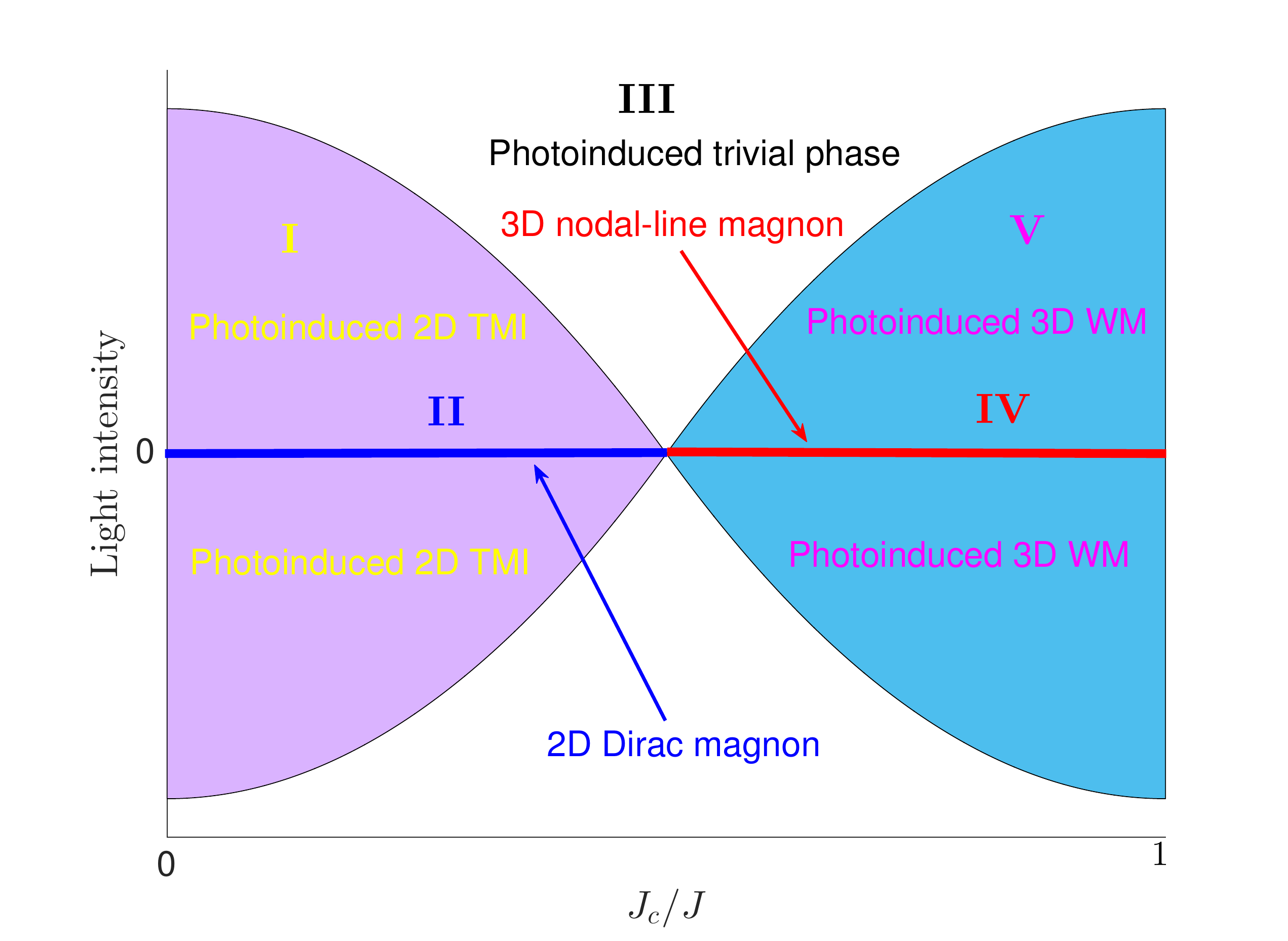}
\caption{Color online. Schematic of the topological phase diagram in periodically driven noncollinear kagome antiferromagnets as a function of the light intensity and the antiferromagnetic interlayer coupling. There are five identified phases. For negligible antiferromagnetic interlayer coupling $J_c/J \ll 1$ there are two phases labeled as $I$ and $II$ corresponding to photoinduced 2D topological magnon insulator (TMI) and undriven 2D Dirac magnon at zero oscillating electric field respectively. For non-negligible antiferromagnetic interlayer coupling $0<J_c/J \leq 1$ there are three phases labeled as $III$, $IV$, and $V$ corresponding to photoinduced 3D trivial magnon insulator, undriven 3D nodal line magnon at zero oscillating electric field, and photoinduced 3D Weyl magnon (WM) respectively.}
\label{phase}
\end{figure}

  Interesting features are expected to manifest in periodically driven insulating antiferromagnets with  zero net magnetization.
Recently, the equilibrium topological aspects of 3D insulating antiferromagnets including Weyl magnons and topological thermal Hall effects have attracted considerable attention \cite{yli,jian,owe,owe1, lau, ylu, owe2, Amook}. They can be induced from 3D noncollinear spin structure by applying an external magnetic field or including an in-plane  Dzyaloshinskii-Moriya  interaction (DM) interaction \cite{dm,dm2}. For the stacked kagome antiferromagnets, the equilibrium Weyl magnon nodes are acoustic and they results from a noncoplanar spin structure with a nonzero scalar spin chirality, which breaks time-reversal symmetry macroscopically \cite{owe}.  The question that remains is whether there is an alternative source to induce  Weyl magnons and thermal Hall effect in 3D insulating noncollinear antiferromagnets.

In this paper, we provide an alternative avenue to induce acoustic Weyl magnon nodes and magnon thermal Hall effect in 3D noncollinear kagome antiferromagnets. We study periodically driven  symmetry-protected 3D in-plane $120^\circ$  spin structure with either positive  or negative vector chirality in the stacked kagome-lattice antiferromagnets.  The photoinduced topological phase diagram of the model is schematically shown in Fig.~\eqref{phase}. In the main paper, we consider regions $IV$ and $V$ when the antiferromagnetic interlayer coupling is not negligible. The 2D  phases for negligible interlayer coupling will be studied in the appendix. The 3D  nodal line magnons and triply-degenerate nodal magnon points are present at zero light intensity in the in-plane $120^\circ$  spin structure. They are protected by crystal and time-reversal symmetries of the kagome lattice \cite{owe}. As the periodic drive is turned on parallel to the  in-plane $120^\circ$  spin structure, we show that its symmetry protection  is broken by the laser field. 

As a result, we obtain non-degenerate magnon quasienergy linear band crossings, which form topological Floquet Weyl magnon nodes with opposite chirality along the $k_x$ direction on the $k_z =\pi$ plane. The photoinduced topological Floquet Weyl magnon nodes are present for both linearly- and circularly-polarized light. However, we show that the photoinduced  thermal Hall effect which characterizes a measurable topological magnon transport property of the system is only present for circularly-polarized light. This means that the thermal Hall transport for linearly-polarized light is topologically trivial. We also show that the photoinduced chiral spin structure is the  canted 3D in-plane $120^\circ$  spin structure  in analogy to the chiral spin structure recently observed in the noncollinear antiferromagnetic Weyl semimetals Mn$_3$Sn\slash Ge \cite{nak, nak1,nay,yang}. Therefore, the current results are different from the equilibrium  Weyl magnons induced by noncoplanar spin structure  with a nonzero scalar spin chirality \cite{owe}.  Our results provide a powerful mechanism for manipulating the intrinsic properties of insulating geometrically frustrated kagome antiferromagnets such as jarosites \cite{men4a}, herbertsmithite \cite{tia}, and Cd-kapellasite \cite{kape}.

 \section{Time-Independent Noncollinear Kagom\'e Antiferromagnets}
 \subsection{Heisenberg spin model}
\begin{figure}
\centering
\includegraphics[width=1\linewidth]{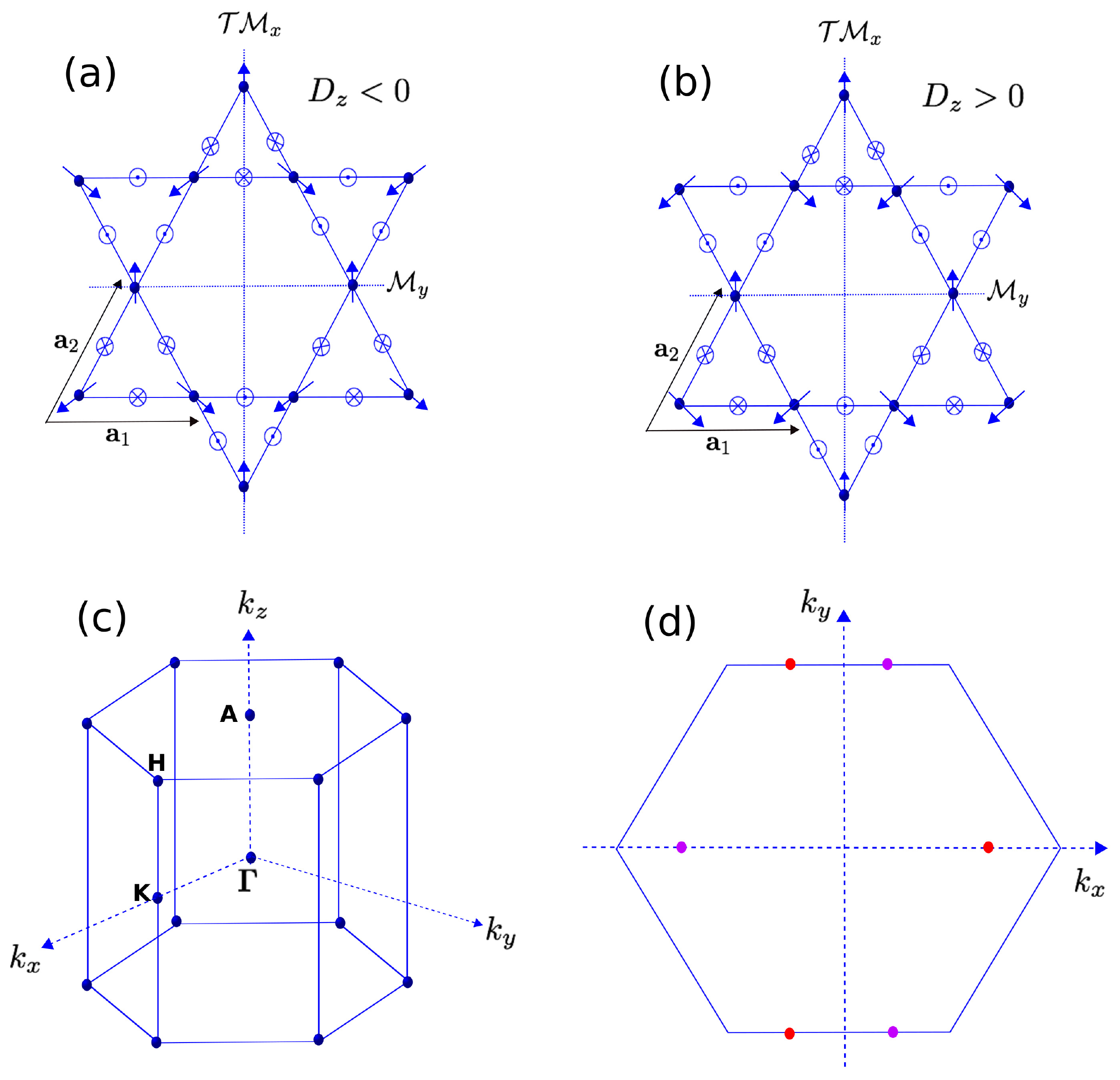}
\caption{Color online. Top panel. Top view of congruently stacked noncollinear kagome antiferromagnets along the (001) direction.  The conventional in-plane $120^\circ$ non-collinear spin configuration is indicated by the blue arrows.  The in-plane unit vectors are ${\bf a}_1=(1,0,0)$ and ${\bf a}_2=(1/2,\sqrt{3}/2,0)$. The unit vector along the stacking direction ${\bf a}_z=(0,0,1)$ is not depicted.  The mirror reflection axes (dotted lines) and the direction of the alternating DM interaction (crossed and dotted circles) are indicated. (a) Positive vector chirality for $D_z<0$. (b) Negative vector chirality for $D_z>0$.    Bottom panel. The Brillouin zone (BZ) of the hexagonal lattice. (c) The bulk BZ with indicated high-symmetry points. (d). The (001) surface BZ with 3 pairs of Floquet Weyl points with opposite chiralities (red and pink dots) along the $k_x$ direction.  }
\label{lattice}
\end{figure}

 We study  stacked  frustrated kagom\'e antiferromagnets governed by the microscopic Heisenberg spin Hamiltonian 
\begin{align}
\mathcal H&=J\sum_{\la ij\ra,\ell} {\vec S}_{i,\ell}\cdot{\vec S}_{j,\ell}+\sum_{\la ij\ra,\ell}{\vec D}_{ij}\cdot{\vec S}_{i,\ell}\times{\vec S}_{j,\ell} \nonumber\\&+J_c\sum_{i,\la \ell \ell^\prime\ra}{\vec S}_{i,\ell}\cdot{\vec S}_{i,\ell^\prime},
\label{ham}
\end{align}
where $i$ and $j$ denote the sites on  the kagom\'e layers, $\ell$ and $\ell^\prime$ label the layers.  The first term corresponds to the nearest-neighbour (NN) antiferromagnetic intralayer Heisenberg interaction. The second term is the out-of-plane (${\vec D}_{ij}=\pm D_z{\hat z}$) DM interaction  due to lack of inversion symmetry between two sites on each kagom\'e  layer. The DM interaction alternates between the triangular plaquettes of the kagom\'e lattice as shown in Figs.~\eqref{lattice}(a) and (b). It also  stabilizes  the conventional 3D in-plane $120^\circ$ non-collinear spin structure. Its sign  determines the vector chirality of the non-collinear spin order \cite{men1}.   The third term  is the NN interlayer antiferromagnetic  Heisenberg interaction  between the kagom\'e  layers. 

\subsection{Symmetry protection of the conventional in-plane $120^\circ$ spin structure}
 \label{sec1}
As explicitly discussed in ref.~\cite{owe}, the conventional 3D in-plane $120^\circ$  spin structure preserves all the symmetries of the kagom\'e lattice. In particular, the combination of time-reversal symmetry (denoted by $\mathcal T$)  and spin rotation denoted by $\mathcal R_z(\pi)$ is a good symmetry, where $\mathcal R_z(\pi)=\text{diag}(-1,-1,1)$ denotes  a $\pi$ spin rotation of the in-plane coplanar spins about the $z$-axis, and `\text{diag}' denotes diagonal elements.  In addition, the system also has three-fold rotation symmetry along the $z$ direction denoted by  $\mathcal C_{3}$. The combination  of mirror reflection symmetry of the kagome plane about the $x$ or $y$ axis  and $\mathcal T$ ({\it i.e.} $\mathcal T\mathcal M_x \mathcal T $ or $\mathcal M_y \mathcal T $ ) is also a symmetry of the conventional in-plane $120^\circ$ non-collinear spin structure. These symmetries are known as the ``effective time-reversal symmetry'' \cite{suzuki, Dgos},  and they lead to protection of nodal-line magnons and triply-degenerate nodal magnon points in the conventional 3D in-plane $120^\circ$  spin structure of the stacked  kagome antiferromagnets. The goal of this paper is to periodically drive these trivial magnon phases to  a topological Floquet Weyl magnon phase without the effects of an external magnetic field or an in-plane DM interaction. 

 \section{Time-Dependent  Noncollinear Kagom\'e Antiferromagnets}
In this section, we will study  regions $IV$ and $V$ in Fig.~\eqref{phase}.  

 \subsection{Theoretical formulation}
 
 The concept of periodically driven magnetic insulators rely on the spin magnetic dipole moments carried by the underlying magnon excitations as previously introduced in quantum ferromagnets \cite{sol0,sol1,sol2}. In other words,   charge-neutral bosonic quasiparticles can interact with an electromagnetic field through their spin magnetic dipole moment. The corresponding time-dependent version of the Aharonov-Casher phase \cite{aha} emerges explicitly from quantum field theory with the Dirac-Pauli Lagrangian  \cite{sol1,sol2}.   We take the spin magnetic dipole moment carried by magnons to be along the in-plane ordering direction $\vec{ \mu}=g\mu_B \hat{x}$. In the presence of  an oscillating electric field  $\vec{E}(\tau)$ propagating along the in-plane $x$ direction, magnons  accumulate the time-dependent version of the Aharonov-Casher phase,  given by
\begin{align}
\Phi_{ij}(\tau)=\mu_m\int_{\vec{r}_i}^{\vec{r}_j} \vec{\Xi}(\tau)\cdot d \vec{\ell},
\label{pha}
\end{align}
 where  $\mu_m= g\mu_B/\hbar c^2$. We have used the notation $\vec{\Xi}(\tau) = \vec E(\tau)\times \hat x$ for brevity, where $\vec E(\tau)=-\partial_\tau{\vec A(\tau)}$ and ${\vec A}(\tau)$ is the time-periodic vector potential. It is important to note that the magnon accumulated time-dependent version of the Aharonov-Casher phase in Eq.~\eqref{pha} is different from that of static time-independent electric field \cite{mei,su} for which the relations in Eqs.~\eqref{eq1} and \eqref{eq3} do not apply. For light propagating along the in-plane $x$ direction  perpendicular to the $y-z$ plane, we choose the time-periodic vector potential such that
 \begin{align}
 \vec{\Xi}(\tau)=\big [0, E_y\sin(\omega \tau), E_z\sin(\omega \tau+\phi)\big],
 \end{align}
where  $\phi$ is the phase difference.   Circularly-polarized electric field corresponds to $\phi=\pi/2$, whereas linearly-polarized electric field corresponds to $\phi=0$ or $\pi$.

\subsection{ Effective spin Hamiltonian}
In the absence of an in-plane DM interaction and an external magnetic field, the  positive and negative vector chiralities  in Figs.~\eqref{lattice}(a) and (b) essentially yield the same magnon dispersions, when an appropriate spin rotation is performed. In fact, both spin configurations in Figs.~\eqref{lattice}(a) and (b) are related by two-fold rotation about the $y$-axis. Therefore, we will restrict our study to the positive vector chirality, and rotate the coordinate axis about the $z$-axis by the spin orientation angles $\theta_{i,\ell}$ 
 
 \begin{align}
 &S_{i,\ell}^x = \cos\theta_{i,\ell} S_{i,\ell}^{\prime x}  -\sin\theta_{i,\ell} S_{i,\ell}^{ \prime y},\\&
 S_{i,\ell}^y = \sin\theta_{i,\ell} S_{i,\ell}^{\prime x}  +\cos\theta_{i,\ell} S_{i,\ell}^{ \prime y},\\&
 S_{i,\ell}^z =  S_{i,\ell}^{\prime z}, 
\label{rot}
\end{align}
where $\theta_{i,\ell}=0, 2\pi/3, -2\pi/3$ for spins on sublattice $A,B,C$ respectively as denoted in Fig.~\eqref{lattice}(a). The prime denotes the rotated coordinate. We    choose the spin quantization axis along the $x$-axis such that $S_{i,\ell}^{\prime\pm} = S_{i,\ell}^{\prime y}\pm i S_{i,\ell}^{\prime z}$ are the raising and lowering operators. 

After substituting the spin transformation into Eq.~\eqref{ham}, the time-dependent phase in Eq.~\eqref{pha} couples to the spin Hamiltonian through the off-diagonal terms such as $\big(S_{i,\ell}^{+\prime}S_{j,\ell}^{-\prime}e^{i\Phi_{ij,\ell}(\tau)} +{\rm H.c.}\big)$ and $\big(S_{i,\ell}^{+\prime}S_{j,\ell}^{+\prime}e^{i\Phi_{ij,\ell}(\tau)} +{\rm H.c.}\big)$ for the intralayer in-plane spin interactions, as well as $\big(S_{i,\ell}^{+\prime}S_{i,\ell^\prime}^{-\prime}e^{i\Phi_{i,\ell\ell^\prime}(\tau)} +{\rm H.c.}\big)$ and $\big(S_{i,\ell}^{+\prime}S_{i,\ell^\prime}^{+\prime}e^{i\Phi_{i,\ell\ell^\prime}(\tau)} +{\rm H.c.}\big)$ for the interlayer spin interaction. In the off-resonant limit when the photon energy $\hbar\omega$ is greater than the energy scale of the undriven system, the effective static Hamiltonian is given by \cite{foot4} $\mathcal H_{eff}\approx \mathcal H_{0}+ \Delta\mathcal H_{eff},$ where $\Delta\mathcal H_{eff}=\big[\mathcal H_{1}, \mathcal H_{-1}\big]/\hbar\omega$  is the photon emission and absorption term and  $\mathcal H_n=\frac{1}{T}\int_0^T d\tau e^{-in\omega \tau} \mathcal H(\tau)$ is the multi-photon Fourier components with period  $T=2\pi/\omega$.  The first term $\mathcal H_{0}$ is the original static spin Hamiltonian with renormalized spin interactions, whereas the second term $\Delta\mathcal H_{eff}$ breaks time-reversal symmetry by  inducing additional  spin  interactions via photon absorption and emission process. For simple spin model  such as the Heisenberg ferromagnet  the effective spin Hamiltonian can be derived explicitly \cite{sol0}. An alternative approach  to magnetic Floquet physics is to consider the charge degree of freedom through the Peierls phase in the Hubbard model, which also maps to a renormalized effective spin model in the off-resonant limit \cite{mat,ste}. This approach, however, does not explicitly  show how the spins couple to the laser electric field in insulating magnets with charge neutral excitation.  Due to the complexity of the present model, the explicit derivation of the effective spin Hamiltonian in the off-resonant limit is very cumbersome. 

\subsection{ Bosonic Bogoliubov-de Gennes Hamiltonian}
Our main objective is to study the effects of light on the magnon band structure of 3D non-collinear antiferrmagnets. Hence, it is advantageous to use linear spin wave theory via the implementation of the linearized Holstein-Primakoff  transformation 
\begin{align}
S_{i,\ell}^{\prime x}= S-a_{i,\ell}^\dagger a_{i,\ell},~ S_{i,\ell}^{\prime +}\approx  \sqrt{2S}a_{i,\ell}=(S_{i,\ell}^{\prime-})^\dg,
\end{align}
 where $a_{i,\ell}^\dagger(a_{i,\ell})$ are the bosonic creation (annihilation) operators. By substituting the Holstein-Primakoff  transformation directly into the resulting time-dependent spin Hamiltonian $\mathcal H(\tau)$, the corresponding time-dependent linear spin-wave Hamiltonian is given by
\begin{widetext}
\begin{align}
\mathcal H(\tau)&= JS\sum_{\la ij\ra,\ell}\Big[t_0(a_{i,\ell}^\dg a_{i,\ell} +a_{j,\ell}^\dg a_{j,\ell}) +t_1 (a_{i,\ell}^\dg a_{j,\ell} e^{i\Phi_{ij,\ell}(\tau)} + {\rm H.c.})  +t_{2}(a_{i,\ell}^\dg a_{j,\ell}^\dg e^{i\Phi_{ij,\ell}(\tau)} +{\rm H.c.})\Big]\nonumber\\& +
J_cS\sum_{i,\la \ell\ell^\prime\ra} \Big[\big(a_{i,\ell}^\dg a_{i,\ell}+a_{i,\ell^\prime}^\dg a_{i,\ell^\prime}\big) +(a_{i,\ell}^\dg a_{i,\ell^\prime}^\dg e^{i\Phi_{i,\ell\ell^\prime}(\tau)} + {\rm H.c.})\Big],
\end{align}
\end{widetext}
where  $t_0=(1+\sqrt{3}D_z/J)/2;~t_1=(1-\sqrt{3}D_z/J)/4;~t_2=(3+\sqrt{3}D_z/J)/4$. The  hopping terms acquire a time-dependent phase factor by virtue of the Peierls substitution as discussed above.  The corresponding time-dependent bosonic Bogoliubov-de Gennes (BdG) Hamiltonian  is given by
  
\begin{align}
\mathcal H(\tau)= \frac{1}{2}\sum_{\vec k}\big(u^\dg(\vec k), u(-\vec k)\big)\cdot
\mathcal H(\vec k,\tau)\cdot
{u(\vec k)\choose u^\dg(-\vec k)},
\end{align}
where  $u^\dg(\vec k)= \big(a_{\bo A}^{\dg},\thinspace a_{\bo B}^{\dg},\thinspace a_{\bo C}^\dg\big)$ is  the basis vector.

 \begin{align}
& \mathcal{H}(\bo,\tau)= 2JS \begin{pmatrix}
  {\mathcal{G}}^d(\bo,\tau)& {\mathcal{G}}^o(\bo,\tau)\\
{\mathcal{G}}^o(\bo,\tau) & {\mathcal{G}}^d(\bo,\tau)
\end{pmatrix}.
\label{eqnr}
\end{align}
The $3\times 3$ matrices are given by 
\begin{align}
&{\mathcal G }^d(\bo,\tau)={\Lambda}^0+t_1{\Lambda}(\vec k_\parallel,\tau),
\\&
{\mathcal G }^o(\bo,\tau)=t_2{\Lambda}(\vec k_\parallel,\tau) + t_c\Lambda^{z}(k_z,\tau),
\end{align}
where $t_c = J_c/J$. 
The  $\Lambda$ matrices are given by
\begin{align}
&{\Lambda}^0 =\big(2t_0 + t_c\big){\rm I}_{3\times 3},~ {\Lambda}^{z}(k_z,\tau)= \cos k_z (\tau){\rm I}_{3\times 3},
\label{mat1}
\end{align}
\begin{align}
{\Lambda}(\vec k^\parallel,\tau) =
\begin{pmatrix}
0& \gamma_{AB}(k^\parallel,\tau)& \gamma_{CA}(k^\parallel,\tau) \\
 \gamma_{AB}^*(k^\parallel,\tau)&0& \gamma_{BC}(k^\parallel,\tau)\\
 \gamma_{CA}^{*}(k^\parallel,\tau)& \gamma_{BC}^{*}(k^\parallel,\tau)&0
\end{pmatrix},
\label{mat1}
\end{align}

where
\begin{align}
&\gamma_{AB}(k^\parallel,\tau) = \cos k_1^\parallel(\tau),
\\&\gamma_{BC}(k^\parallel,\tau) = \cos k_2^\parallel(\tau),
\\&\gamma_{CA}(k^\parallel,\tau) = \cos k_3^\parallel(\tau),
\end{align}
with $k_i(\tau) = k_i + \mu_m \Xi(\tau)$ and $k_i^\parallel=\vec k\cdot{\vec  a}_i$, with  ${\vec a}_1=a\big(\hat x/2 +\sqrt{3}\hat y/2\big) $,    ${\vec a}_2 = a\hat x$, and ${\vec a}_3={\vec a}_1-{\vec a}_2$.

\begin{figure*}
\centering
\includegraphics[width=.9\linewidth]{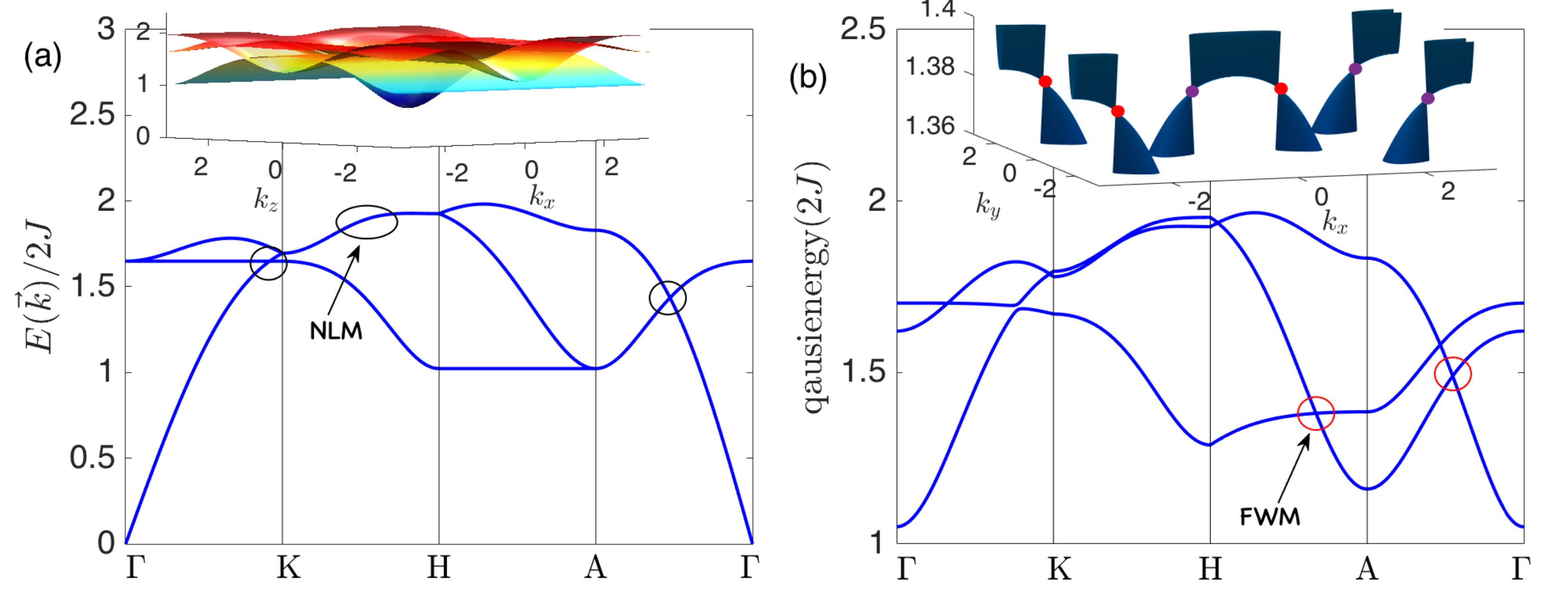}
\caption{Color online. (a) The magnon band structures of undriven 3D noncollinear kagome antiferromagnet showing a coexistence of 3D nodal-line magnons and triply-degenerate nodal magnon points for  $S=1/2$, $D_z/J=0.2$, $J_c/J=0.5$, $\mathcal {E}_y=\mathcal {E}_z =0$. The black circles denote the nodal-line magnon (NLM) nodes. The magnon band crossing along $A$--$\Gamma$ line form triply-degenerate nodal magnon points.  Inset shows the 3D magnon bands on the $k_y=0$ plane.  (b) The magnon quasienergy bands in the periodically driven 3D noncollinear kagome antiferromagnet for $S=1/2$, $D_z/J=0.2$, $J_c/J=0.5$, $\mathcal {E}_y=\mathcal {E}_z =1$,  $\phi = \pi/2$ or $\phi =0$, and $\omega/J = 10$. The red circles denote the photoinduced Floquet Weyl magnon (FWM) nodes at the acoustic branch. Inset shows the 3D topological Floquet Weyl magnon nodes on the $k_z= \pi$ plane. There are 3 pairs of Floquet Weyl magnon nodes with opposite chiralities on the $k_z= \pi$ plane located at $\vec k_{Q_1} = \big[\pm\cos^{-1}(Q_1), 0, \pi\big]$ and $\vec k_{Q_2} = \big[\pm\cos^{-1}(Q_2), \pm \pi/3, \pi\big]$. In addition, please note that the incident light breaks U(1) spin rotation and lifts the Goldstone modes at the $\Gamma$ point.}
\label{FWM_band}
\end{figure*}

\begin{figure*}
\centering
\includegraphics[width=0.9\linewidth]{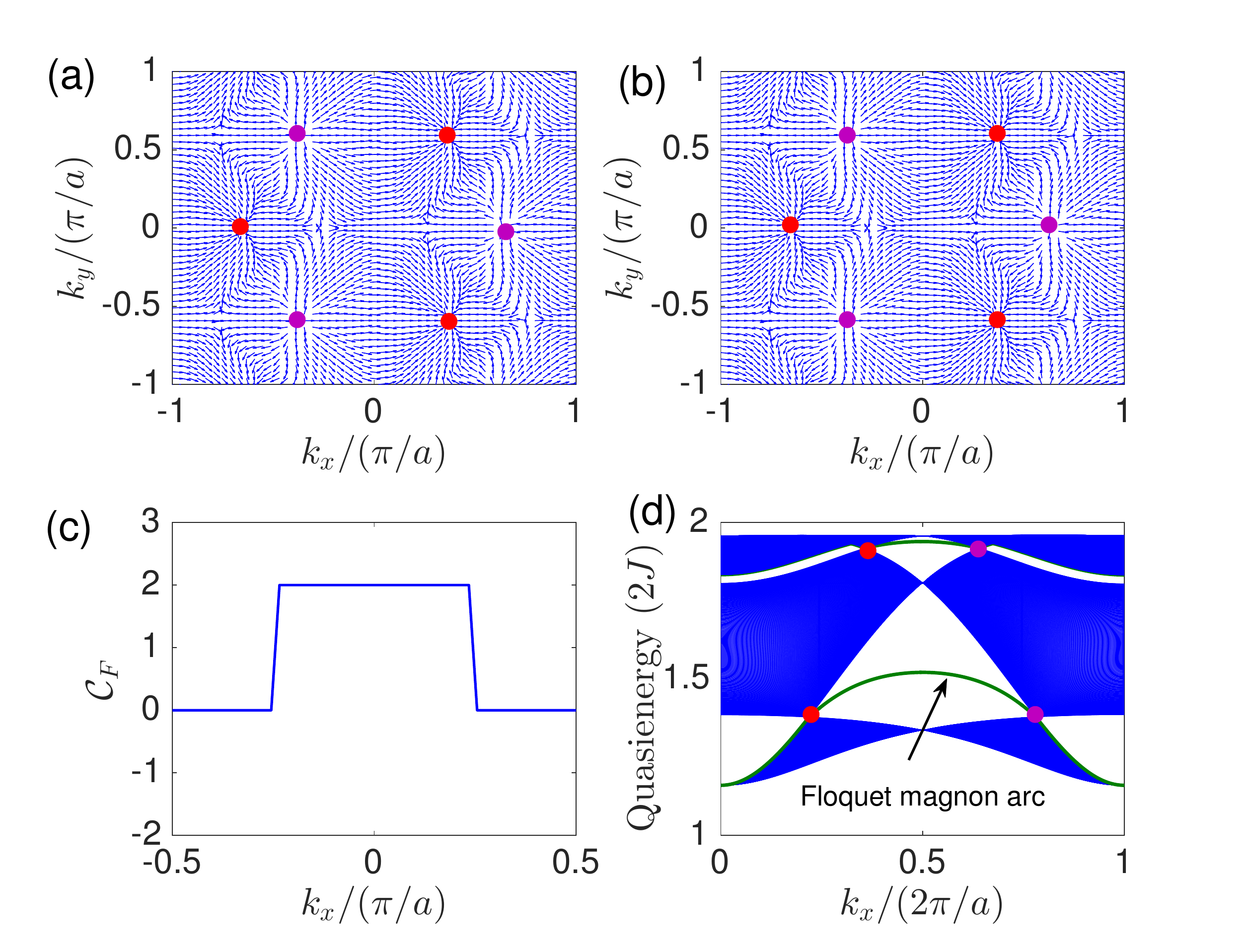}
\caption{Color online. Top panel. Photoinduced Berry curvature monopoles of the acoustic topological Floquet Weyl magnon nodes  on the $k_z = \pi$ plane for (a) linearly-polarized light $\phi =0$ and (b) circularly-polarized light $\phi =\pi/2$. Bottom panel. (c) The Floquet Chern number of the acoustic magnon quasienergy band as function of $k_x$ for circularly-polarized light $\phi =\pi/2$. (d) (010)-surface magnon quasienergy band structure along a line connecting a pair of Floquet Weyl magnon nodes  with opposite chirality  for circularly-polarized light $\phi =\pi/2$ at $k_z=\pi$. The parameters are  $S=1/2$, $D_z/J=0.2$, $J_c/J=0.5$, $\mathcal {E}_y=\mathcal {E}_z =1$ and $\omega/J=10$.}
\label{BC}
\end{figure*}
\subsection{Bosonic Floquet-Bloch theory}

The Floquet theory is a powerful mechanism to study periodically driven quantum systems.  In this formalism  we can  transform the time-dependent bosonic BdG Hamiltonian $\mathcal H(\bo,\tau)$ into a static effective Hamiltonian governed by  the Floquet bosonic BdG Hamiltonian.  To do this we expand the time-dependent bosonic BdG Hamiltonian as
\begin{align}
\mathcal H(\vec{k},\tau)=\sum_{n=-\infty}^{\infty} e^{in\omega \tau}\mathcal H_{n}(\vec{k}),
\end{align}
where $\mathcal H_{n}(\vec{k})=\frac{1}{T}\int_{0}^T e^{-in\omega \tau}\mathcal H(\vec{k}, \tau) d\tau=\mathcal H_{-n}^\dg(\vec{k})$ are the Fourier components. The corresponding  eigenvectors can be written as $\ket{\psi_{n^\prime}(\vec{k}, \tau)}=e^{-i \epsilon_{n^\prime}(\vec{k}) \tau}\ket{\chi_{n^\prime}(\vec{k}, \tau)},$
where  $\ket{\chi_{n^\prime}(\vec{k}, \tau)}=\ket{\chi_{n^\prime}(\vec{k}, \tau+T)}=\sum_{n} e^{in\omega \tau}\ket{\chi_{n^\prime}^n(\vec{k})}$ is the time-periodic Floquet-Bloch wave function of magnons and $\epsilon_{n^\prime}(\vec{k})$ are the magnon quasienergy bands. We define the Floquet operator  as $\mathcal H^F(\vec{k},\tau)=\mathcal H(\vec{k},\tau)-i\partial_\tau$. 
The corresponding eigenvalue equation is of the form 
\begin{align}
\sum_m \big[\mathcal H_{n-m}(\vec{ k}) + m\omega\delta_{n,m}\big]\chi_{{n^\prime}}^m(\vec{k})= \epsilon_{n^\prime}(\vec{k})\chi_{{n^\prime}}^n(\vec{k}).
\end{align}
The Fourier components of the Floquet bosonic BdG Hamiltonian are given by

 \begin{align}
& \mathcal{H}_q(\bo)= 2JS \begin{pmatrix}
  {\mathcal{G}}_q^d(\bo)& {\mathcal{G}}_q^o(\bo)\\
{\mathcal{G}}_q^o(\bo) & \mathcal{G}_q^d(\bo)
\end{pmatrix},
\end{align}
where $q = n-m \in \mathbb{Z}$. The $3\times 3$ matrices are given by 
\begin{align}
&\mathcal{G}_q^d(\bo)=\Lambda^0\delta_{q,0}+t_1{\Lambda}_q(\vec k_\parallel),
\\&
\mathcal G_q^o(\bo)=t_2{\Lambda}_q(\vec k_\parallel) + t_c\Lambda_q^{z}(k_z).
\end{align}
The  $\Lambda_q$ matrices are given by
\begin{align}
&\Lambda^{z}_q(k_z)= \frac{1}{2}\lb \tilde \Lambda^{z}_q(k_z)+ \tilde \Lambda^{z*}_{-q}(k_z)\rb{\rm I}_{3\times 3},\\&
\Lambda_q(\vec k^\parallel) = \frac{1}{2}\lb \tilde \Lambda_{q}(\vec k^\parallel) + \tilde \Lambda_{-q}^*(\vec k^\parallel)\rb,
\label{eq24}
\end{align}
where
\begin{align}
\tilde \Lambda^{z}_q(k_z) =\mathcal J_{q}(\mathcal E_z) e^{ik_z} e^{iq \phi},
\end{align}
\begin{align}
\tilde {\Lambda}_{q}(\vec k^\parallel) =
\begin{pmatrix}
0& f_{q, AB}& f_{q,CA}\\
 f_{q,AB}^* &0& f_{q, BC}\\
 f_{q,CA}^{*} & f_{q,BC}^{*}&0
\end{pmatrix},
\label{eq25}
\end{align}
with
\begin{align}
&f_{q,AB} =  \mathcal J_{q}(\sqrt{3}\mathcal E_y/2)e^{ik_1^\parallel},
\\&f_{q,BC} =  \delta_{q,0}e^{ik_2^\parallel},\\&  f_{q,CA}  = \mathcal J_{q}(\sqrt{3}\mathcal E_y/2)e^{ik_3^\parallel}.
\end{align}
Here  $\mathcal J_{q}(x)$ is the Bessel function of order $q$.  The intensity of light in the magnonic Floquet formalism  is characterized by the dimensionless quantity  $ \mathcal E_i$ given by Eq.~\eqref{eq1}.


 \begin{figure*}
\centering
\includegraphics[width=1\linewidth]{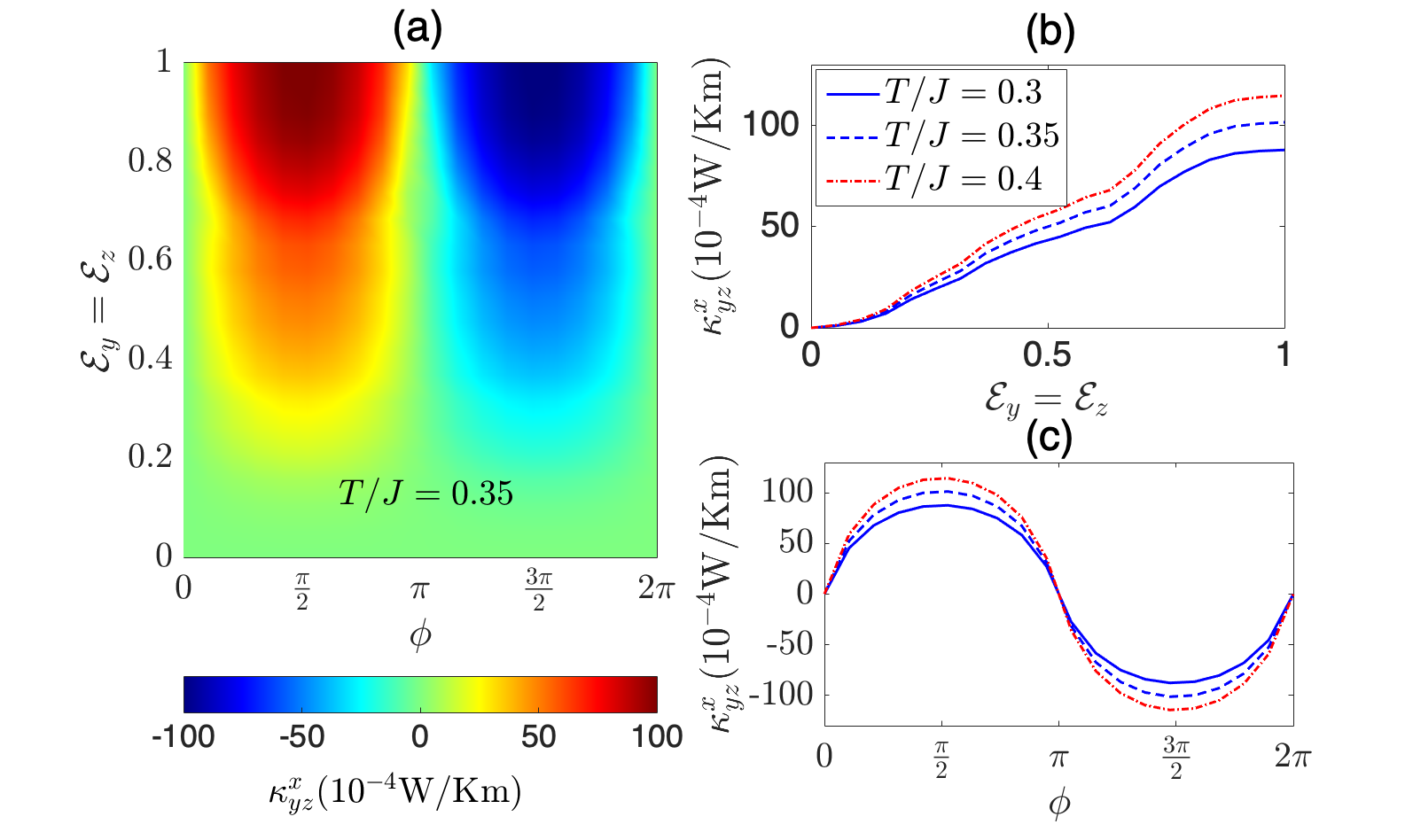}
\caption{Color online. Photoinduced Floquet thermal Hall conductivity $\kappa_{yz}^x$ in the (a) $\phi$ -- $\mathcal{E}$ plane for $T/J=0.35$, (b) as a function of  light intensity $\mathcal {E}_y=\mathcal {E}_z$ for $\phi =\pi/2$, and (c)  as a function of phase difference $\phi$ for $\mathcal {E}_y=\mathcal {E}_z=1$. The other parameters are  $S=1/2$, $D_z/J=0.2$, $J_c/J=0.5$ and $\omega/J=10$. }
\label{THE}
\end{figure*}

 \section{High-frequency limit}

We will study the off-resonant limit $\hbar\omega\gg J, D_z, J_c$, when the system can be described by an effective time-independent  Hamiltonian \cite{we4,gru}, given by
\begin{align}
\mathcal H_{eff}(\vec k)&=\mathcal H_{0}(\vec k)-\frac{1}{\hbar\omega}\big[\mathcal H_{-1}(\vec k), \mathcal H_{1}(\vec k)\big].
\label{effHam}
\end{align}
As we mentioned above, in terms of the original spin operators, the first term is the original static spin model with renormalized interactions, whereas the second term  breaks time-reversal symmetry by  inducing additional spin  interactions.  The effective Floquet bosonic BdG Hamiltonian $\mathcal H_{eff}(\vec k)$  can be  diagonalized numerically using the generalized Bogoliubov transformation. To do this, we make a linear  transformation $\psi(\bo)= \mathcal{P}(\bo) \mathcal{M}(\bo)$, 
where $\mathcal{P}(\bo)$ is a $2N\times 2N$ paraunitary matrix  defined as
\begin{align}
& \mathcal{P}(\bo)= \begin{pmatrix}
  u_\bo& -v_\bo^* \\
-v_\bo&u_\bo^*\\  
 \end{pmatrix},
\end{align} 
where $u_\bo$ and $v_\bo$ are  $N\times N$ matrices that satisfy  the relation
\bea 
|u_\bo|^2-|v_\bo|^2={\rm I}_{N\times N}.
\eea 
The  quasiparticle operators are given by $ \mathcal{M}^\dg(\bo)= \big(m^\dg(\bo),\thinspace m(-\bo)\big)$ with $ m^\dg(\bo)=(b_{\bo A}^{\dg}, b_{\bo B}^{\dg}, b_{\bo C}^{\dg})$. 
The paraunitary operator $\mathcal{P}(\bo)$ satisfies the relations
\begin{align}
&\mathcal{P}^\dg(\bo) \mathcal{H}_{eff}(\bo) \mathcal{P}(\bo)=\mathscr{E}(\bo),\quad \mathcal{P}^\dg(\bo) {\tau}_3 \mathcal{P}(\bo)= {\tau}_3,
\label{eqna}
\end{align}
where
\begin{align}
\mathscr{E}(\bo) =
\begin{pmatrix}
\epsilon_{n}(\bo) & 0\\
0&\epsilon_{n}(-\bo)
\end{pmatrix},\quad \tau_3 =
\begin{pmatrix}
{\rm I}_{N\times N} & 0\\
0&-{\rm I}_{N\times N}
\end{pmatrix}.
\end{align}
 Using  Eq.~\eqref{eqna}, we see that the Hamiltonian to be diagonalized  is $\mathcal{H}_{eff}^B(\bo)=  \tau_3\mathcal{H}_{eff}(\bo),$ whose eigenvalues are given by $ \tau_3 \mathscr{E}(\bo)$ and the columns of $\mathcal P(\bo)$ are the corresponding eigenvectors.
 
  \subsection{Photoinduced Floquet Weyl magnon bands}

In Fig.~\eqref{FWM_band}(a) we have shown the undriven ({\it i.e.} $\mathcal E_y =\mathcal E_z=0$) magnon bands of 3D noncollinear spin structure in the kagome antiferromagnets along the BZ paths depicted in Fig.~\eqref{lattice}(c). The system possesses  nodal line magnon  at  the $A$ point and along the lines $\Gamma$--$K$ and $K$--$H$. There also exists a triply-degenerate nodal magnon points along the line $A$--$\Gamma$ where three magnon branches linearly cross each other. The details of these magnon nodal line phases have been previously discussed \cite{owe}, and they are protected by the crystal symmetry of the kagome lattice as discussed in Sec.~\eqref{sec1}. As the periodic drive is turned on, we expect  the symmetry protection of the nodal line magnons to be broken by the light intensity. 

Indeed, in Fig.~\eqref{FWM_band}(b) we can see that the  nodal line magnons at the $A$ point is split into six Floquet Weyl magnon (Floquet Weyl magnon)  nodes (only two are independent) located along the $k_x$ direction at $\vec k_{Q_1} = \big[\pm\cos^{-1}(Q_1), 0, \pi\big]$ and $\vec k_{Q_2} = \big[\pm\cos^{-1}(Q_2), \pm \pi/3, \pi\big]$.  Due to the complexity of this model we could not find an explicit analytical form for $Q_1$ and $Q_2$. There are 3 pairs of Floquet Weyl magnon nodes with opposite chiralities on the $k_z=\pi$ plane as indicated in Fig.~\eqref{lattice}(d). The triply-degenerate nodal magnon points along the $A$--$\Gamma$ line is also split into two Floquet Weyl magnon nodes  along the $k_z$ direction. In contrast to 2D systems, we find that both linearly-polarized lights ($\phi = 0$ and $\phi = \pi$) and circularly-polarized lights ($\phi = \pi/2$) break  time-reversal symmetry and induce Floquet Weyl magnon nodes in the 3D system. The reason is because the $y$ and $z$ directions are not symmetric in the present system. However, as we will show in the subsequent section, the Floquet magnon thermal transports for circularly-polarized light is topologically nontrivial, whereas those for linearly-polarized light is  topologically trivial. We also note that the incident light breaks U(1) spin rotation and lifts the Goldstone modes at the $\Gamma$ point as shown in Fig.~\eqref{FWM_band}(b).
 
\subsection{Photoinduced Berry curvature}
 Usually, the analysis of the band structure is not sufficient to proof that a non-degenerate linear band crossing in 3D system forms topological Weyl nodes.  The existence of topological Weyl nodes can be reinforced by computing the Berry curvature of the linear band crossing. Topological Weyl nodes act as the source and sink of the Berry curvature.  This implies that a single Weyl node can be considered as the monopole of the Berry curvature. Using the  paraunitary operator,  we can define the photoinduced Floquet Berry curvature as

\begin{align}
\Omega_{\alpha\beta,n}^{\gamma}(\bo)=-2{\rm Im}\sum_{m\neq n}\frac{\big[ \braket{\mathcal{P}_{\bo n}|\hat v_{\alpha}|\mathcal{P}_{\bo m}}\braket{\mathcal{P}_{\bo m}|\hat v_{\beta}|\mathcal{P}_{\bo n}}\big]}{\big[ \epsilon_{m}(\bo)-\epsilon_{n}(\bo)\big]^2},
\label{chern2}
\end{align}
where $\hat v_{\alpha}=\partial \mathcal{H}^B_{eff}(\bo)/\partial k_{\alpha}$ defines the photoinduced velocity operators with $\alpha,\beta,\gamma=x,y,z$ and $n$ labels the Floquet magnon branches. The photoinduced Berry curvature can be considered as a 3-pseudo-vector pointing along  the $\gamma$ directions perpendicular to both the $\alpha$ and $\beta$ directions. It diverges at the photoinduced Floquet Weyl magnon nodes as can been seen from the denominator of Eq.~\eqref{chern2}. 

 In Fig.~\eqref{BC}(a) and Fig.~\eqref{BC}(b) we have shown the photoinduced  Berry curvature  $\Omega_{xy,1}^{z}(\bo)$ in the $k_z=\pi$ plane for the acoustic Floquet magnon band. Indeed, we can see that the topological Floquet Weyl magnon nodes  at $\vec k_{Q_1}$ and $\vec k_{Q_2}$ are the monopoles and anti-monopoles of the Floquet Berry curvature for both linearly- and circularly-polarized lights.  
 
Along the line $H$--$A$ there are two independent Floquet Weyl points located at $\pm k_x^c$. For the 2D planes for fixed $k_x \neq \pm k_x^c$ the system is gapped and can be considered as a Floquet Chern insulator. Hence, we can define the Floquet Chern number for fixed $k_x$ as
\begin{align}
\mathcal C_{n}^F(k_x) = \frac{1}{2\pi}\int_{BZ} dk_ydk_z~\Omega_{yz,n}^{x}(k_x,k_y,k_z).
\end{align}
In Fig.~\eqref{BC}(c) we have shown the plot of the Floquet Chern number of the acoustic magnon quasienergy branch as a function of $k_x$ for circularly-polarized light $\phi=\pi/2$. We can see that $\mathcal C^F=0$ for planes with  $k_x<-k_x^c$ or $k_x>k_x^c$, while $\mathcal C^F=2$ for planes with  $-k_x^c<k_x<k_x^c$. This implies that the magnon quasienergy band crossing between the planes with  $\mathcal C^F=2$ and  $\mathcal C^F=0$ are indeed Floquet Weyl points.

One of the hallmarks of equilibrium Weyl semimetals is the formation of Fermi arcs which connect the surface projection of the Weyl points in the BZ. Indeed, as shown in Fig.~\eqref{BC}(d)  we can see that the (010)-projected Floquet Weyl magnon points  with opposite chirality  are connected by  Floquet magnon arc surface states. These hallmark properties of equilibrium Weyl semimetals confirm that the non-degenerate magnon quasienergy linear band crossings in the present system are indeed topological Floquet Weyl magnon points.

\subsection{Photoinduced thermal Hall effect}

Another hallmark of equilibrium Weyl semimetals is the appearance of the anomalous Hall effect. For magnon quasiparticles with no charge, the corresponding transport property  is the anomalous thermal Hall effect in the presence of a temperature gradient \cite{alex, alex1a, alex1b, alex1c, alex1d}. In the nonequilibrium system, it is customary to assume the limit in which the occupation function of the quasiparticle exciations is close to equilibrium. Therefore, we can  write the photoinduced thermal Hall conductivity  in the usual form \cite{alex1b}
\begin{align}
\kappa_{\alpha\beta}^{\gamma} =- k_BT\int_{BZ} \frac{d\bo}{(2\pi)^3}~ \sum_{n=1}^N c_2\lb f_n^B\rb\Omega_{\alpha\beta,n}^{\gamma}(\bo),
\label{thm}
\end{align}
where   $ f_n^B=\big( e^{\epsilon_{n}(\bo)/k_BT}-1\big)^{-1}$ is the Bose occupation function close to equilibrium, $k_B$ the Boltzmann constant which we will  set to unity, $T$ is the temperature  and $ c_2(x)=(1+x)\lb \ln \frac{1+x}{x}\rb^2-(\ln x)^2-2\text{Li}_2(-x)$ weight function with $\text{Li}_2(x)$ being the  dilogarithm.

In Fig.~\eqref{THE} we have shown the trends of the photoinduced thermal Hall conductivity $\kappa_{yz}^{x}$  in the $\phi$ -- $\mathcal{E}$ plane at $T/J=0.35$ (a), as a function of $\mathcal{E}_y=\mathcal{E}_z$ (b), and as a function of $\phi$ (c) for $T/J =0.3,0.35,0.4$. In Fig.~\eqref{THE}(b), we can see that $\kappa_{yz}^{x}$ vanishes in the limit of undriven system $\mathcal{E}_y=\mathcal{E}_z=0$. This is due to the presence of time-reversal symmetry in the undriven  in-plane $120^\circ$  spin structure, which leads to zero Berry curvature. Although  the photoinduced Berry curvature of the Floquet Weyl points is nonzero for linearly-polarized light $\phi=0$ or $\phi=\pi$, we can see in Fig.~\eqref{THE}(c) that $\kappa_{yz}^{x}$  vanishes at $\phi=0$ and $\phi =\pi$.  Therefore, the Floquet magnon thermal transports  for linearly-polarized light is topologically trivial.
 
\subsection{Photoinduced chiral spin structure}

We would like to address the nature of the photoinduced spin structure that gives rise to the topological  magnon properties in this system.  It is evident that  in the periodically driven 3D non-collinear kagome antiferromagnets, there is no photoinduced noncoplanar spin structure with nonzero scalar spin chirality. This is because light propagates parallel to the in-plane $120^\circ$  spin structure (i.e. along the in-plane $x$ direction). Therefore, the topological Floquet Weyl magnons can only originate from a photoinduced canted in-plane $120^\circ$ chiral  spin structure with no finite scalar spin chirality, but time-reversal symmetry is broken through the second term in Eq.~\eqref{effHam}.  We note that the intrinsic equilibrium form of this canted in-plane $120^\circ$ chiral spin structure was observed recently  in the electronic stacked kagome noncollinear antiferromagnets Mn$_3$Sn\slash Ge \cite{nak, nak1,nay,yang}, which are also antiferromagnetic Weyl semimetals. Therefore, the mechanism that gives rise to the topological Floquet Weyl magnons is different from that of equilibrium  Weyl magnon nodes induced by noncoplanar chiral spin structure \cite{owe}.

\section{Conclusion and Outlook}
We have shown that laser-irradiated noncollinear stacked kagome antiferromagnets with a  symmetry-protected three-dimensional (3D) in-plane $120^\circ$  spin structure can be tuned to a topological Floquet Weyl magnon semimetal.  They arise when  the incident light propagates parallel to the  in-plane $120^\circ$  spin structure. We showed that the  topological Floquet Weyl magnon semimetal in the periodically driven noncollinear stacked kagome antiferromagnets originate from a photoinduced canted 3D in-plane $120^\circ$ chiral spin structure, whose intrinsic equilibrium  form was recently observed in the electronic noncollinear stacked kagome antiferromagnets Mn$_3$Sn\slash Ge \cite{nak, nak1,nay,yang}.  We also studied the experimental measurable photoinduced thermal Hall effect due to the Berry curvature of the Floquet Weyl magnon nodes.  We believe  that the current results are within experimental reach and can be accessible with the current terahertz frequency using ultrafast terahertz spectroscopy. The current results also pave the way for manipulating the intrinsic properties of geometrically frustrated kagome antiferromagnets.  It is believed that topological antiferromagnets have potential technological applications in spintronics \cite{pea,magn}, because they have zero net spin magnetization which makes their magnetism  externally invisible and insensitive to external magnetic fields.

Most studies in Floquet topological systems customarily assume that the quasienergy levels of the Floquet Hamiltonian are close to the equilibrium system  \cite{foot3,foot4,lin, fot,foot5}. Therefore, the properties of equilibrium topological systems can be applied to  Floquet topological systems. However, Floquet topological systems are inherently out of equilibrium and thus the question about the non-equilibrium  distribution function of the quasiparticles becomes important. In the present case, the distribution function of magnon enters observable quantities  such as the thermal Hall effect. In the non-equilibrium system, the distribution function of magnon will depend on how the periodic drive is switched on \cite{deh,deh1}. Therefore, the topological properties of the non-equilibrium system will be completely different from that of  equilibrium system. To study the  non-equilibrium system, one can consider an open system in which magnon couples to external reservoir of phonons. Alternatively, one can consider the dynamics of the system in a quantum quench when the periodic drive is suddenly switched on. However, a complete analysis of this study is beyond the purview of this paper. We plan to address this problem in future work as we continue to develop the study of Floquet topological magnons.

\section*{ACKNOWLEDGEMENTS}
Research at Perimeter Institute is supported by the Government of Canada through Industry Canada and by the Province of Ontario through the Ministry of Research and Innovation.

\section*{APPENDIX A. PHOTOINDUCED 2D TOPOLOGICAL MAGNON INSULATOR}

\subsection{Time-independent spin model}

In this appendix, we will study phases $I$ and $II$ in Fig.~\eqref{phase}. In these phases  the interlayer coupling can be neglected and thus the Hamiltonian reduces to a strictly 2D system.   We will consider the 2D strained  Heisenberg antiferromagnetic spin Hamiltonian given by
\begin{align}
\mathcal H&=\sum_{\la ij\ra} J_{ij} {\vec S}_{i}\cdot{\vec S}_{j}+\sum_{\la ij\ra}{\vec D}_{ij}\cdot{\vec S}_{i}\times{\vec S}_{j},
\label{appenham}
\end{align}
where  $J_{ij}=J$ on the diagonal bonds and $J_{ij}=J^\prime=J\delta$ on the horizontal bonds with $\delta \neq 1$ being the strain or lattice distortion. The DM interaction is still out-of-plane, {\it i.e.}  ${\vec D}_{ij}=-D_z{\hat z}$. For $D_z=0$ the classical ground state  yields the strained induced canting angle
\bea 
\varphi= \arccos\lb -\frac{1}{2}\delta \rb.
\eea 
Note that $\varphi\neq 120^\circ$ for $\delta \neq 1$. The limiting case $\delta\to 0$ maps to a bipartite square-lattice with collinear magnetic order, whereas  $\delta\to \infty$ maps to a decoupled antiferromagnetic spin chain. The canted coplanar spin structure is stabilized for $\delta >1/2$, but it also has an extensive degeneracy as in the case of ideal kagome Heisenberg antiferromagnet at $\delta =1$. However, the extensive degeneracy is lifted by a finite DM interaction  $D_z\neq 0$ and thus a unique ground state is selected. 
 \begin{figure*}
\centering
\includegraphics[width=0.9\linewidth]{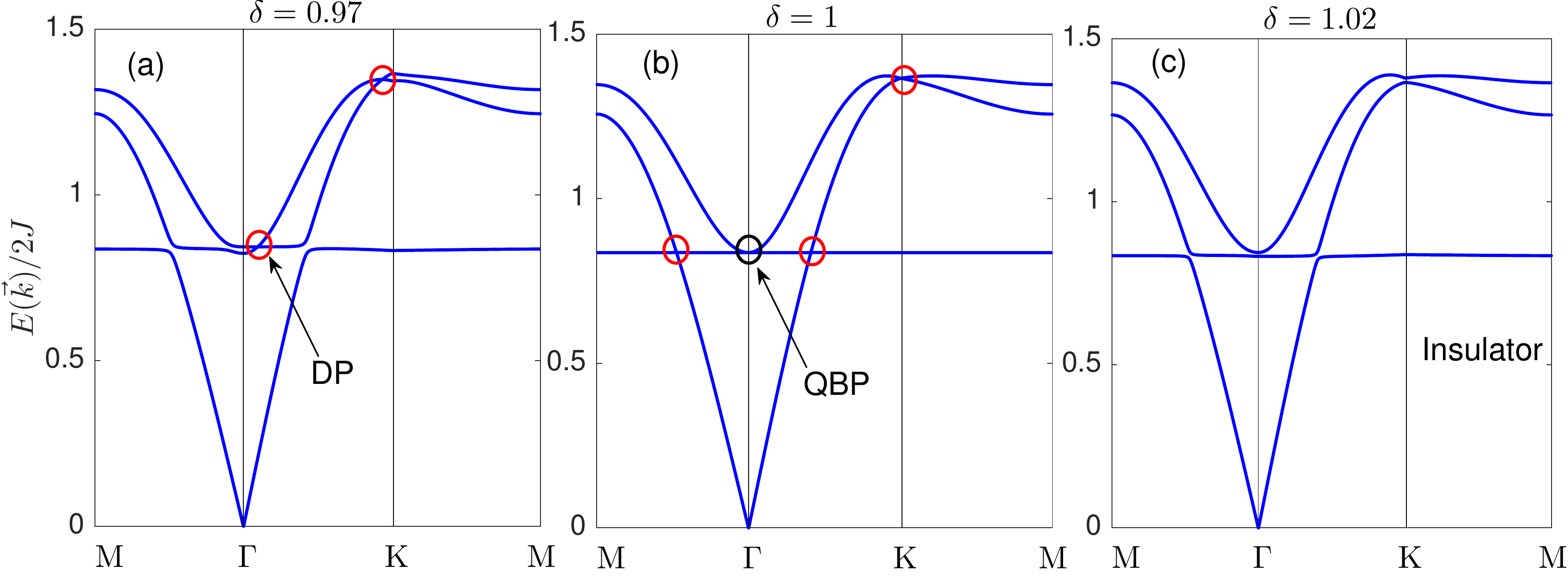}
\caption{Color online. Strained-induced magnon phase transition of undriven  2D kagome antiferromagnets for $D_z/J=0.2$ and $\mathcal{E}_x=\mathcal{E}_y =0$. (a) $\delta =0.97$ with Dirac point (DP) along the $\Gamma$-$K$ line. (b) $\delta =1$ isotropic limit with quadratic band crossing point (QBP) at $\Gamma$. (c) $\delta =1.02$ with no magnon band crossing point.}
\label{2D_band}
\end{figure*}
 \begin{figure*}
\centering
\includegraphics[width=0.9\linewidth]{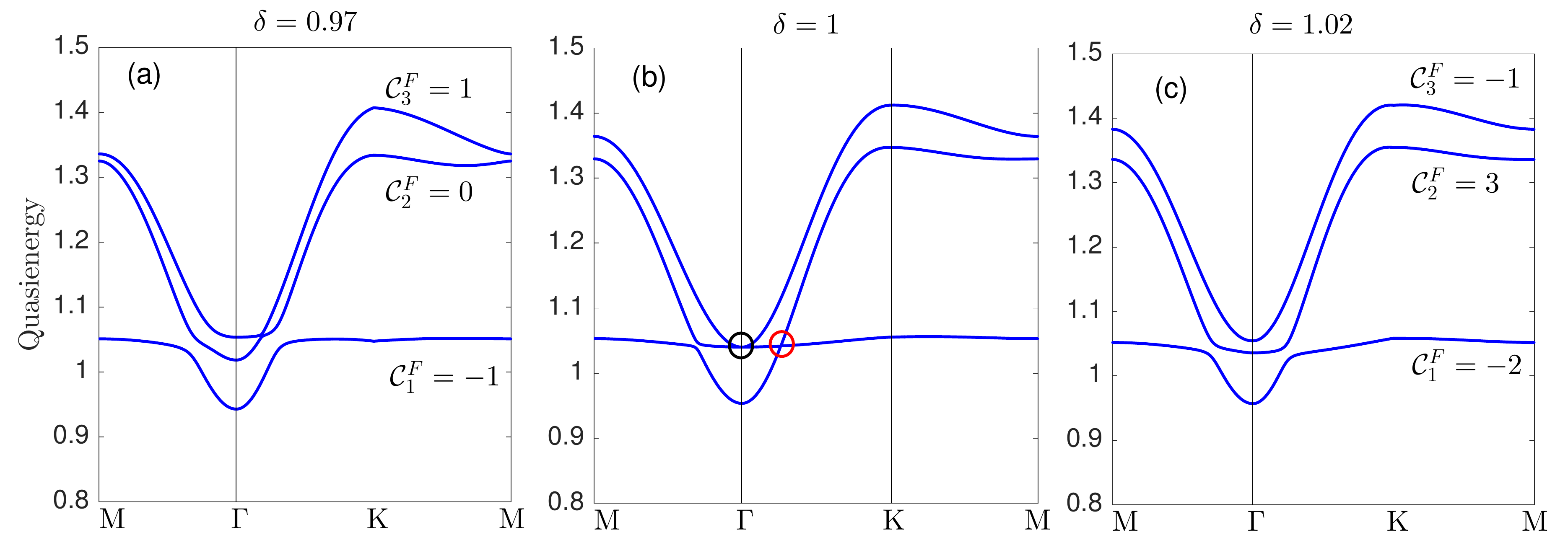}
\caption{Color online. Floquet strained-induced topological magnon phase transition of periodically driven 2D kagome antiferromagnets for $D_z/J=0.2$ and $\mathcal{E}_x=\mathcal{E}_y =1$ and $\phi =\pi/2$. (a) $\delta =0.97$ with Floquet magnon Chern numbers $\mathcal C^F=\big(-1,0,-1\big)$. (b) $\delta =1$ phase transition point. (c) $\delta =1.02$ with Floquet magnon Chern numbers $\mathcal C^F=\big(-2,3,-1\big)$. }
\label{2D_F_band}
\end{figure*}

\subsection{Time-dependent bosonic model}
We will study the magnon excitations of the noncollinear system for $\delta >1/2$ in the presence of an oscillating electric field propagating along the $z$-direction perpendicular to the $x$-$y$ plane, which is given by
\begin{align}
 \vec{\Xi}(\tau)=\big [E_y\sin\omega \tau, E_x\sin(\omega \tau+\phi), 0\big],
 \end{align}
where $\vec{\Xi}(\tau) = \vec E(\tau)\times \hat z$.

 Following the same procedure outlined above, we find that the Fourier components of the Floquet bosonic BdG Hamiltonian are given by

 \begin{align}
& \mathcal{H}_q(\bo)= 2JS \begin{pmatrix}
  {\mathcal{G}}_q^d(\bo)& {\mathcal{G}}_q^o(\bo)\\
{\mathcal{G}}_q^o(\bo) & \mathcal{G}_q^d(\bo)
\end{pmatrix}.
\end{align}
The $3\times 3$ matrices is this case have a different form given by 
\begin{align}
&\mathcal{G}_q^d(\bo)=\Lambda^0\delta_{q,0}+{\Lambda}_q^{(1)}(\vec k_\parallel),\quad 
\mathcal G_q^o(\bo)={\Lambda}_q^{(2)}(\vec k_\parallel),
\end{align}
where
 \begin{figure*}
\centering
\includegraphics[width=0.9\linewidth]{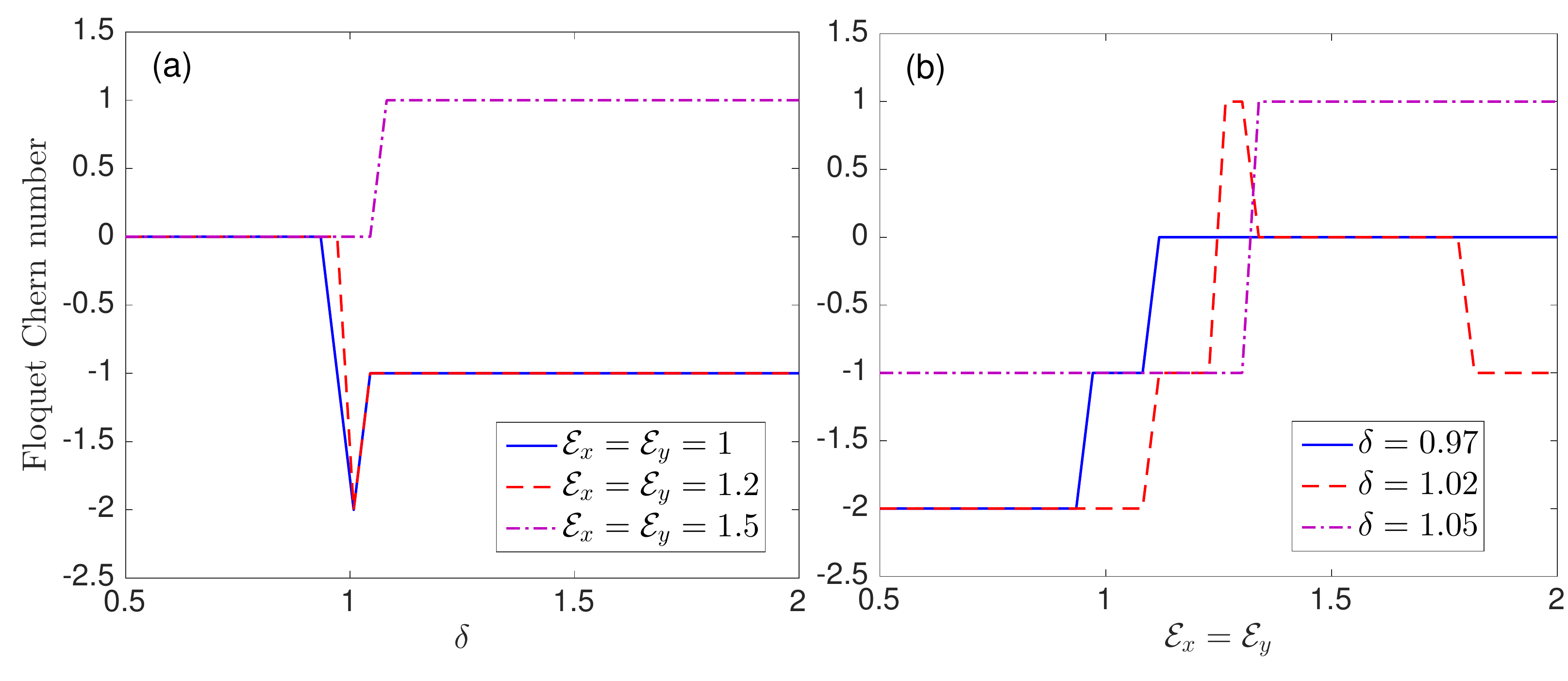}
\caption{Color online. Plot of Floquet Chern number of the lowest quasienergy magnon band as a function of $(a)$ $\delta$ and (b) $\mathcal{E}_x=\mathcal{E}_y$, for $\phi=\pi/2$.}
\label{F_ChernN}
\end{figure*}

\begin{align}
\Lambda^0 = {\rm diag}\lb \xi_{AA}, \xi_{BB}, \xi_{CC}\rb,
\end{align}
with the parameter given by
\begin{align}
&\xi_{AA} = -2\big(\cos\varphi - D_J \sin\varphi\big),\\&
\xi_{BB}= \xi_{CC} =-\delta\big(\cos2\varphi +D_J\sin2\varphi\big) +\xi_{AA}/2.
\end{align}
We have used the notation $D_J = D_z/J$ for brevity. The  $\Lambda_q^{(i)}$ matrices are given by
\begin{align}
\Lambda_q^{(i)}(\vec k^\parallel) = \frac{1}{2}\lb \tilde \Lambda_{q}^{(i)}(\vec k^\parallel) + \big[\tilde \Lambda_{-q}^{(i)}\big]^*(\vec k^\parallel)\rb,
\label{eq24}
\end{align}
where
\begin{align}
\tilde {\Lambda}_{q}^{(i)}(\vec k^\parallel) =
\begin{pmatrix}
0& a_if_{q, AB}& a_if_{q,CA}\\
 a_if_{q,AB}^* &0& \tilde{a_i}f_{q, BC}\\
 a_if_{q,CA}^{*} & \tilde{a_i}f_{q,BC}^{*}&0
\end{pmatrix}.
\label{eq25}
\end{align}

The constant $a_i$ factors are given by
\begin{align}
&a_1= \frac{1}{2}\lb 1+\cos\varphi -D_J\sin\varphi\rb,\\&
\tilde{a}_1= \delta\lb\cos 2\varphi +\sin^2\varphi + D_J\sin\varphi\cos\varphi\rb,\\&
a_2= \frac{1}{2}\lb 1-\cos\varphi +D_J\sin\varphi\rb,\\&
\tilde{a}_2= \delta\sin^2\varphi -D_J\sin\varphi\cos\varphi.
\end{align}
The  $f$ functions are given by
\begin{align}
&f_{q,AB} =  \mathcal J_{q}(\mathcal E_+)e^{iq\Psi_+}e^{ik_1^\parallel},
\\&f_{q,BC} =  \mathcal J_{q}(\mathcal E_x)e^{ik_2^\parallel},\\&  f_{q,CA}  =  \mathcal J_{q}(\mathcal E_-)e^{-iq\Psi_-}e^{ik_3^\parallel},
\end{align}
where $\mathcal  J_n(x)$ is the Bessel function of order $n$.
\begin{align}
 &\mathcal{E}_{\pm}=\frac{1}{2}\sqrt{3\mathcal{E}_x^2+\mathcal{E}_y^2\pm 2\sqrt{3}\mathcal{E}_x \mathcal{E}_y\cos(\phi)},\\& \Psi_{\pm}=\arctan\lb \frac{\sqrt{3}\mathcal{E}_x\sin(\phi)}{\mathcal{E}_y\pm \sqrt{3}\mathcal{E}_x\cos(\phi)}\rb.
 \end{align}
 
\subsection{Photoinduced Floquet topological magnon phase transition } 

We will now study the topological magnon phase transition associated with the 2D strained kagome antiferromagnets in the presence of circularly-polarized electric field.  Let us first consider the undriven system.  In Fig.~\ref{2D_band} we have shown the magnon bands of the undriven 2D strained kagome antiferromagnets for varying $\delta$. We can see that the Dirac magnon ``semimetal'' in the regime $0<\delta\leq 1$ can be turned to a trivial insulator for $\delta >1$. As shown Fig.~\ref{2D_F_band}, we can see that circularly-polarized electric field induces a Floquet topological magnon phase transition  across the topological phase boundary at the isotropic point $\delta =1$.  To investigate the complete photoinduced Floquet topological magnon phase transition, we define the Floquet Chern number of quasienergy magnon bands as
\begin{align}
\mathcal C_{n}^F = \frac{1}{2\pi}\int_{BZ} d^2k~\Omega_{xy,n}^z(\vec k). 
\end{align}
In Fig.~\ref{F_ChernN} we have shown the evolution of the lowest Floquet magnon Chern number as a function of $\delta$ for varying amplitudes (a) and as a function of $\mathcal E_x=\mathcal E_y$ for  varying $\delta$ (b). In both cases we have considered circularly-polarized electric field $\phi = \pi/2$. We note that the Floquet Chern number vanishes for linearly polarized light in the 2D systems. We can see that the lowest Floquet magnon Chern number varies in the range $\mathcal C_{1}^F= (-2, -1,0, 1)$ which indicates a photoinduced Floquet topological magnon phase transition in the driven 2D strained kagome antiferromagnets.

\end{document}